\pdfoutput=1
\documentclass[a4paper,american,citeautoscript,floatfix,pdftex,superscriptaddress,twoside,%
aps,prb,round,%
longbibliography,%
reprint,twocolumn,final
]{revtex4-1}%
\usepackage{amsfonts,amsmath,amssymb}
\usepackage{commath}
\usepackage[T1]{fontenc}
\usepackage{graphicx}
\usepackage[utf8]{inputenc}
\usepackage{microtype}
\usepackage[obeyFinal,textsize=footnotesize]{todonotes}
\usepackage{xspace}
\usepackage{hyperref, hypernat}
\usepackage[todos]{CMImacros}

\hypersetup{citebordercolor=white,linkbordercolor=white,urlbordercolor=white}
\makeatletter
\renewcommand\NAT@biblabelnum[1]{(#1)}
\makeatother

\newcommand{\cfeldesy}{\affiliation{Center for Free-Electron Laser Science, Deutsches
      Elektronen-Synchrotron DESY, Notkestraße 85, 22607 Hamburg, Germany}}%
\newcommand{\uhhcui}{\affiliation{The Hamburg Center for Ultrafast Imaging, Universität Hamburg,
      Luruper Chaussee 149, 22761 Hamburg, Germany}}%
\newcommand{\uhhphys}{\affiliation{Department of Physics, Universität Hamburg, Luruper Chaussee 149,
      22761 Hamburg, Germany}}%
\newcommand{\ayemail}{\email[]{andrey.yachmenev@cfel.de}}%
\newcommand{\cmiweb}{\homepage{https://www.controlled-molecule-imaging.org}}%

\begin{document}
\title{Coherent Control of the Rotation Axis of Molecular Superrotors}
\author{A. Owens}\cfeldesy\uhhcui%
\author{A. Yachmenev}\ayemail\cmiweb\cfeldesy\uhhcui%
\author{J. Küpper}\cfeldesy\uhhcui\uhhphys%
\date{\today}
\begin{abstract}\noindent%
   The control of ultrafast molecular rotational motion has benefited from the development of
   innovative techniques in strong-field laser physics. Here, we theoretically demonstrate a novel
   type of coherent control by inducing rotation of an asymmetric-top molecule about two different
   molecular axes. An optical centrifuge is applied to the hydrogen sulfide (H$_2$S) molecule to
   create a molecular superrotor, an object performing ultrafast rotation about a well-defined axis.
   Using two distinct pulse envelopes for the optical centrifuge, we show that H$_2$S can be excited
   along separate pathways of rotational states. This leads to stable rotation about two entirely
   different molecular axes while ensuring rotation is about the propagation direction of the
   centrifuge, \ie, the laboratory-fixed $Z$-axis. The presented scheme to control the angular
   momentum alignment of a molecule will, for instance, be useful in studies of molecule-molecule or
   molecule-surface scattering, especially due to the large amounts of energy associated with
   superrotors, which can even be controlled by changing the duration of the optical centrifuge
   pulse. \\
   \begin{center}
      \includegraphics[width=5cm]{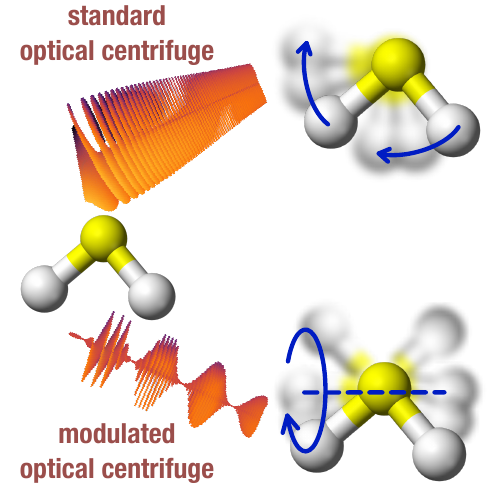}
   \end{center}
\end{abstract}
\maketitle

A long-standing goal of physics and chemistry is to control the motion of molecules. Tailored
external fields offer a means to achieve this~\cite{Lemeshko:MP111:1648}, from alignment and
orientation where molecules can be spatially fixed in the laboratory
frame~\cite{Stapelfeldt:RMP75:543, Holmegaard:PRL102:023001, Ghafur:NatPhys5:289}, to the creation
of unidirectional rotating molecular ensembles that possess oriented angular
momentum~\cite{Fleischer:NJP11:105039, Kitano:PRL103:223002, Zhdanovich:PRL107:243004}. This latter
application, associated with controlled rotation, has benefited from a number of clever techniques
to prepare molecules in highly excited rotational states~\cite{Ohshima:IRPC29:619,
   Milner:ACP159:395}. Perhaps the most efficient of these is the optical
centrifuge~\cite{Karczmarek:PRL82:3420, Villeneuve:PRL85:542}, which is a strong, nonresonant,
linearly-polarized laser pulse that performs accelerated rotation about the direction of
propagation. The centrifuge field aligns the most polarizable molecular axis, trapping and forcing
the molecule to follow the rotating laser polarization. In doing so, molecules can be adiabatically
spun into extremely high angular momentum states, creating molecular ``superrotors'', objects with
narrow, well-defined rotational wavepackets that are resistant to collisions and
reorientation~\cite{Yuan:PNAS108:6872, Milner:PRL113:043005, Khodorkovsky:NatComm6:7791}.

The majority of studies on superrotors have involved diatomic and linear triatomic systems. Less
explored are asymmetric-top molecules, which have complicated rotational spectra and thus more
complex dynamics. Recently, an optical centrifuge was able to induce field-free planar alignment of
SO$_2$ with the O--S--O plane confined to the plane of the laser polarization long after the
centrifuge pulse had ended~\cite{Korobenko:PRL116:183001}. This demonstrated that the rotational
dynamics of an asymmetric-top could be reduced to that of a simple linear rotor, which was later
confirmed theoretically~\cite{Omiste:PRA97:023407}, and also highlighted the potential of the
optical centrifuge as a tool for alignment and orientation. Further applications involving
asymmetric-top molecules in an optical centrifuge include a novel approach to manipulate enantiomers
in the gas phase through enantioselective orientation~\cite{Tutunnikov:JPCL9:1105}.

\begin{figure}
   \includegraphics[width=\linewidth]{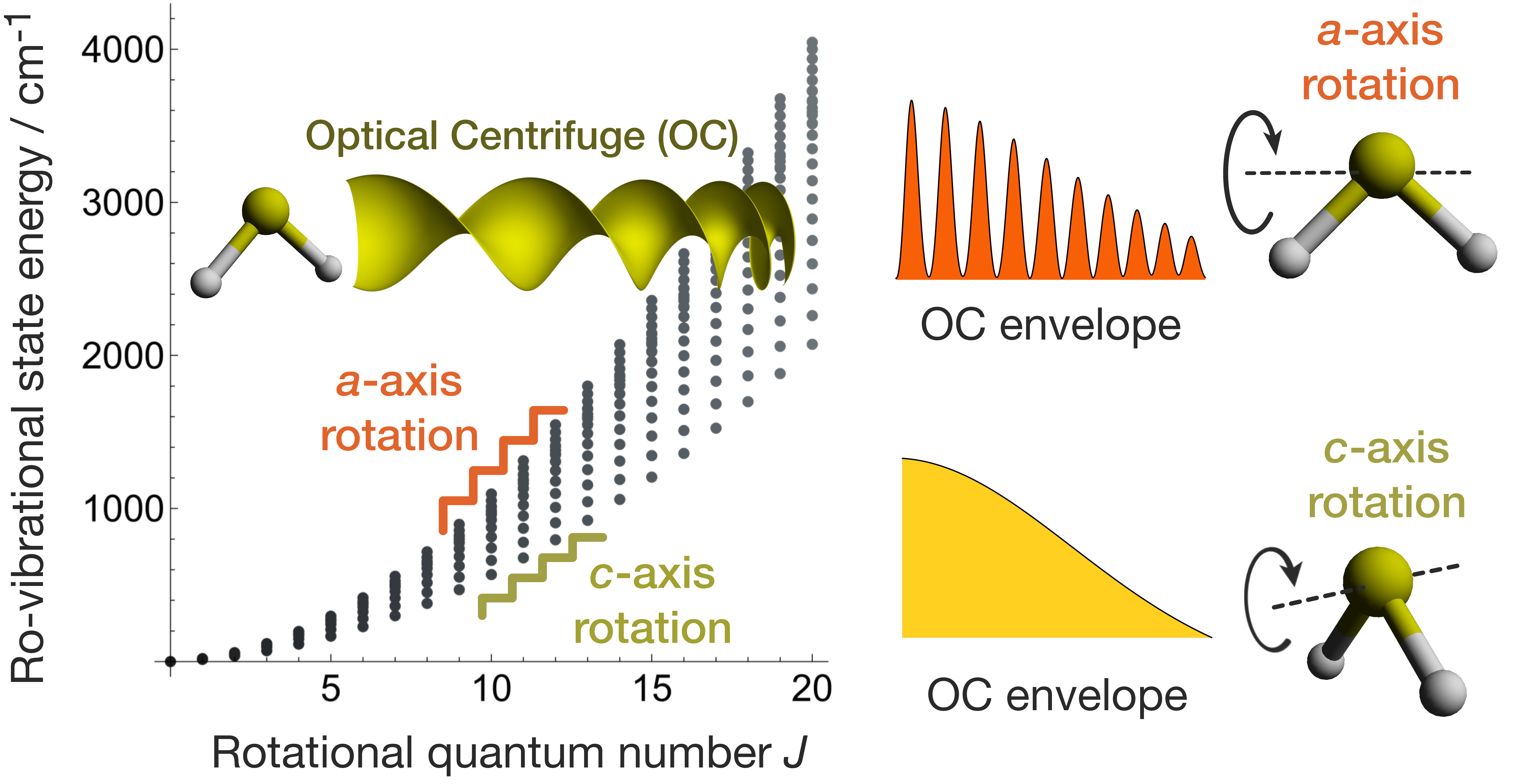}
   \caption{Controlling the molecular axis of rotation in H$_2$S superrotors using an optical
      centrifuge. A Gaussian pulse envelope produces rotation about the principal $c$-axis and
      confinement of the H--S--H plane to the laboratory-fixed $XY$ plane of the laser polarization.
      Using an optimized sinc function combined with a Gaussian profile for the pulse envelope, the
      molecule can be excited along a different pathway of rotational states leading to rotation
      about the principal $a$-axis.}
   \label{fig:scheme}
\end{figure}
Here, we extend the possible types of rotational control over asymmetric-top molecules using
nonresonant laser pulses. Our scheme is outlined in \autoref{fig:scheme}: An optical centrifuge is
applied to the hydrogen sulfide (H$_2$S) molecule to create a molecular superrotor. Normally,
rotation takes place about the principal $c$-axis and the H--S--H plane is confined to the plane of
the rotating laser polarization. Modifying the pulse envelope of the optical centrifuge, stable
rotation about an entirely different molecular axis, namely the principal $a$-axis, can be obtained
while still ensuring rotation is about the propagation direction of the centrifuge, \ie, the
laboratory-fixed $Z$-axis. This is achieved by exciting the molecule through a different pathway of
rotational states, illustrating a novel type of coherent control involving ultrafast rotation.
Schemes to change the angular momentum alignment of a molecule have been demonstrated, for example,
to alter the outcome of H$_2$ molecule-surface collisions~\cite{Godsi:NatComm8:15357}. Our results
show that such stereodynamics studies are possible for asymmetric-top molecular superrotors, which
possess large amounts of energy that, furthermore, can be controlled by adjusting the duration of
the centrifuge pulse.

To realize the scheme in \autoref{fig:scheme}, we utilize a robust, quantum mechanical approach that
treats all major electronic, nuclear motion, and external field effects with high accuracy.
Calculations are geared toward experimentally realizable conditions, \eg, laser parameters and
timescales. The stationary rovibrational energies and eigenfunctions of H$_2$S up to $J=40$, where
$J$ is the quantum number of the total angular momentum operator, were computed from direct
numerical variational calculations~\cite{Yurchenko:JMS245:126, Yachmenev:JCP143:014105,
   Yurchenko:JCTC13:4368} with a highly accurate, three-dimensional potential energy
surface~\cite{Azzam:MNRAS460:4063}. External field effects, \ie, the matrix elements of the electric
polarizability tensor, were evaluated in the basis of rovibrational states. The \emph{ab initio}
polarizability surface was generated using the coupled cluster method, CCSD (coupled cluster with
all single and double excitations), with the augmented correlation-consistent triple-zeta basis set,
aug-cc-pVTZ(+d for S) in the frozen core approximation, and then fitted with a suitable, symmetrized
analytic representation. Electronic structure calculations employed the quantum chemistry package
DALTON~\cite{Aidas:WIRE4:269}. For the quantum dynamics simulations, which used the computer program
RichMol~\cite{Owens:JCP148:124102}, the time-dependent wavefunction was built from a superposition
of field-free rovibrational states and the time-dependent coefficients determined by numerical
solution of the time-dependent Schrödinger equation using the split-operator method.

The electric field of the optical centrifuge is highly oscillating and off-resonant, therefore only
the polarizability term is required in the field interaction potential,
\begin{equation}
   V(t) = -\frac{1}{2}\sum_{A,B=X,Y} \epsilon_{A}^\text{oc}(t)\epsilon_{B}^\text{oc}(t)\alpha_{AB},
\end{equation}
where $\alpha_{AB}$ are the components of the effective electric polarizability in the
laboratory-fixed $XYZ$-axis system. The optical centrifuge ${\boldsymbol\epsilon}^\text{oc}(t)$ was
applied along the laboratory-fixed $Z$-axis for a duration of $t=300$~ps and was represented by the
expression,
\begin{equation}\label{eq:oc}
   \hspace*{-1.5mm}%
   {\boldsymbol\epsilon}^\text{oc}(t) = f(t)\,\epsilon_0\cos(\omega t)\left\{
      \mathbf{e}_X\cos(\beta t^2) ; \mathbf{e}_Y\sin(\beta t^2) \right\}
\end{equation}
with the peak amplitude of the field \mbox{$\epsilon_0=8.6\times10^{13}~\text{W/cm}^2$}, the
acceleration of circular rotation $\beta=(2\pi c)^2\cdot 4$~cm$^{-2}$, the carrier frequency of the
field $\omega=c/(2\pi{}\cdot800~\text{nm})$, and the pulse envelope function
\begin{equation}
   f(t)=\exp(-t^2/2\sigma^2),
\end{equation}
where $\sigma=140$~ps~\cite{Tutunnikov:JPCL9:1105}. Simulations were started in the rovibrational
ground state ($J=k=m=0$) which is valid for temperatures of $T\le3$~K. The quantum numbers $k$ and
$m$ correspond to the projections, in units of $\hbar$, of $J$ onto the molecule-fixed $z$-axis and
laboratory-fixed $Z$-axis, respectively. Low temperatures are necessary in asymmetric-top molecules
to ensure a narrow distribution of states in the initial wavepacket and, consequently, stable
rotational dynamics.

\begin{figure}
   \includegraphics[width=\linewidth]{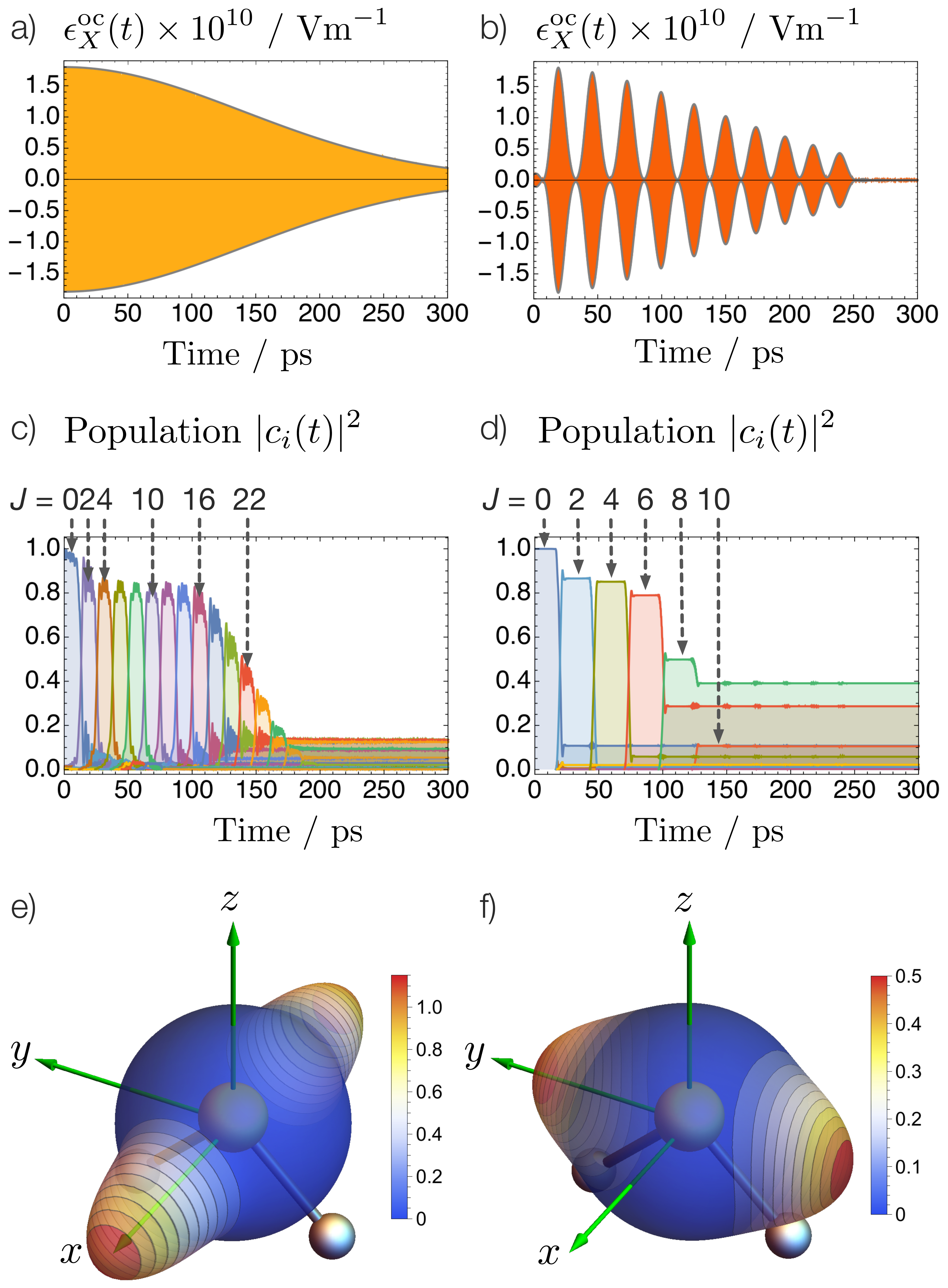}
   \caption{Wavepacket population $|c_i(t)|^2$ and rotational probability density function
      $P(\theta,\chi)$ of H$_2$S in an optical centrifuge with a pulse envelope modeled by a
      Gaussian function (left panels), or with an optimized sinc function combined with a Gaussian
      (right panels). The $X$-projection of the electric field is shown at the top, the wavepacket
      population as a function of time in the middle, and $P(\theta,\chi,t=290~\text{ps})$ at the
      bottom; $x,y,z$ refer to the molecule-fixed axis system.}
   \label{fig:pulse+probability}
\end{figure}
The time evolution of the external electric field for a Gaussian pulse envelope and the
corresponding wavepacket populations of H$_2$S are plotted in \autoref[a
and~c]{fig:pulse+probability}, along with the rotational probability density function
$P(\theta,\chi)=\int\dif{V}\dif\phi\,\psi(t)^{\ast}\psi(t)\sin\theta$ at the end of the centrifuge
pulse in \autoref[e]{fig:pulse+probability}. The probability density function illustrates the
orientation of the molecule with respect to the possible axes of rotation, where the Euler angles
are denoted $\theta,\chi,\phi$, the volume element $\dif{V}$ is associated with the three
vibrational degrees of freedom, and $\psi(t)$ is the wavepacket. H$_2$S is primarily excited through
$\Delta{J}=2$, $\Delta{m}=-2$ rotational Raman transitions and proceeds along a pathway of lowest
energy rotational states within each $J$, \ie, those corresponding to rotation about the molecular
$c$-axis. At $t=102$~ps, the wavepacket is dominated by the $J=16,m=-16$ state at 1360~cm$^{-1}$,
but the maximum single-state population starts to rapidly decline after this time. This is due to
the decreasing intensity of the centrifuge field, suggesting that the optical centrifuge pulse
should be truncated between $t\approx100$--$120$~ps to ensure H$_2$S is efficiently released in a
high $J$ state. The rotational density plot shows two clear ``islands'', which correspond to
rotation about the principal $c$-axis in either a clockwise or anticlockwise direction. The H--S--H
plane is confined to the $XY$ plane of the laser polarization which is consistent with previous
studies of SO$_2$ in an optical centrifuge~\cite{Korobenko:PRL116:183001, Omiste:PRA97:023407}.

To change the axis of rotation, H$_2$S has to be excited along a different pathway of rotational
states. This can be achieved using an optical centrifuge with a modified pulse envelope such that
the field intensity is at a maximum only when the angular frequency of the centrifuge is in
resonance with the desired transition frequency between two states, thus enabling population
transfer. The intensity is then minimized as the angular frequency increases and sweeps through
undesired transitions before being maximized again at a later resonant frequency. The time points
$t_i^\text{res}$ at which the angular frequency of the optical centrifuge in \eqref{eq:oc} reaches
resonance with the desired rotational transition with frequency $\nu_i^\text{res}$ can be evaluated
as $t_i^\text{res}=\nu_i^\text{res}/(4\beta)$. Now, to force H$_2$S to rotate about the principal
$a$-axis the rotational excitation must be guided through the upper energy rotational state within
each $(2J+1)$ multiplet. The respective $J\rightarrow J+2$ transition wavenumbers are 58.4, 138.4,
220.0, 300.1, and 377.6~\invcm, corresponding to frequencies of 1750.8, 4149.1, 6595.4, 8996.7, and
11\,320.1~GHz, for $J=0, 2, 4, 6, 8$, which result in resonant times of
$t_i^\text{res}=19.4,45.9,73.0,99.6,125.3$~ps, respectively, for the centrifuge acceleration
parameter $\beta=(2\pi c)^2\cdot 4$~cm$^{-2}$. The pulse profile function is modeled by a sum of
sinc functions centered at the different resonant times $t_i^\text{res}$ weighted with a Gaussian
envelope function, that is
\begin{equation}
   f(t) = \sum_{i} \left(\frac{\sin((t-t_i^\text{res})/\sigma_s)}{(t-t_i^\text{res})/\sigma_s}\right)^2 \exp(-t^2/2\sigma^2),
\end{equation}
where $\sigma_s=4$~ps and $\sigma=140$~ps. The resulting pulse profile function is plotted in
\autoref[b]{fig:pulse+probability} together with the results of the simulation in \autoref[d and
f]{fig:pulse+probability}. At $t=102$~ps, the wavepacket is dominated by the $J=8,m=-8$ rotational
state at 717~\invcm, which corresponds to rotation about the principal $a$-axis, indicated by the
two shifted islands on the probability density plot. Since this is a lower $J$ state, the rotational
density is broader and not as pronounced as in the first scenario. Because $m=-J$, the axis of
rotation is along the laboratory-fixed $Z$-axis, \ie, the propagation direction of the optical
centrifuge. Essentially, the molecule has assumed a perpendicular orientation compared to the case
of the standard centrifuge excitation shown in \autoref[e]{fig:pulse+probability}. This can be
analyzed by looking at the quantum numbers $k_a$ and $k_c$ corresponding to the projection of the
total angular momentum onto the molecule-fixed $a$ and $c$-axis, respectively. For standard
centrifuge excitation $k_c\approx{}J$, while for the modified pulse envelope excitation
$k_a\approx{}J$. Note that the sinc function pulse envelope could also be optimized for standard
centrifuge-type excitation, but this does not yield any major improvements in population transfer.

\begin{figure}
   \includegraphics[width=\linewidth]{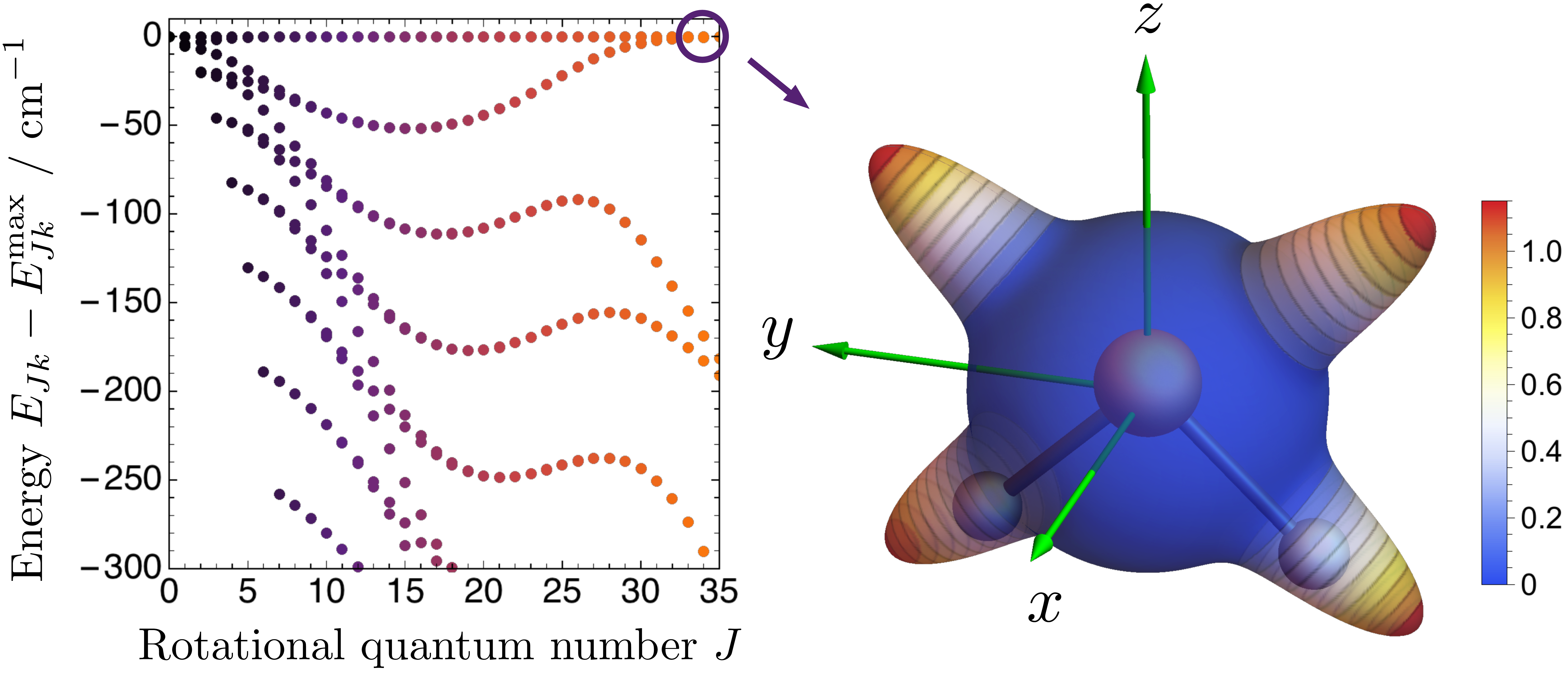}
   \caption{Rotational energy level clustering in the ground vibrational state of H$_2$S. In the
      left panel, the energy difference $E_{Jk}-E_{Jk}^{\text{max}}$ has been plotted for each
      rotational state, with energy $E_{Jk}$, relative to the maximum energy $E_{Jk}^{\text{max}}$
      in its $J$ multiplet. As $J$ increases, after a certain critical $J$ value the energy
      difference between energy levels starts to decrease and rotational cluster states
      form~\cite{Pyka:MolPhys70:547, Kozin:JMolSpec163:483}. In the right panel, the rotational
      probability density function $P(\theta,\chi)$ is illustrated for the top cluster state at
      $J=35$, which shows four possible axes of rotation coinciding with the S--H bonds. Here, $xyz$
      refers to the molecule-fixed axis system.}
   \label{fig:clustering}
\end{figure}

Since H$_2$S is a relatively light, fast rotor, efficient rotational excitation to much higher $J$
would require centrifuge pulses of several-hundred-picosecond duration~\cite{Trippel:MP111:1738}.
Interestingly, H$_2$S displays rotational energy level clustering for large
$J$~\cite{Pyka:MolPhys70:547, Kozin:JMolSpec163:483}, a characteristic effect of polyatomic
molecules with local mode vibrations~\cite{Jensen:WCMS2:494}. After a certain critical $J$ value,
which for H$_2$S is around $J_c\approx15$, the rotational energy levels form fourfold degenerate
clusters as shown in \autoref{fig:clustering} for the ground vibrational state.
Assuming population can be transferred along the correct pathway of rotational states, as H$_2$S
climbs the rotational ladder from $J\approx 16$ to $J\approx 30$, the axis of rotation gradually
moves from the principal $a$-axis to an axis coinciding with one of the S--H bonds, as shown in the
probability density plot of \autoref{fig:clustering}. These so-called cluster states also have the
unique property of being chiral~\cite{Bunker:JMolSpec228:640}, where rotation in either a clockwise
or anticlockwise direction corresponds to the two rotating enantiomers of a dynamically chiral
system~\cite{Owens:RIC:inprep}. However, while it could offer intriguing opportunities for
controlling chirality, clustering is not a necessary condition for the coherent control method
proposed here, which is applicable to molecules of arbitrary structure.

In conclusion, we have shown that asymmetric-top molecules can be made to rotate in a stable manner
about two entirely different molecular axes. This was achieved for H$_2$S using optical centrifuges
with different pulse envelopes, tailored to excite the molecule though distinct pathways of
rotational states. The proposed scheme is relatively robust and there is some tolerance for
shot-to-shot fluctuations in the chirp rate $\beta$ and the sinc function width parameter
$\sigma_s$, e.g., at least $\pm 1\%$ for the values used in this study. Furthermore, for larger
values of $\sigma_s$ and/or smaller values of $\beta$, the robustness against such fluctuations will
increase. Lower field intensities, although resulting in less efficient rotational excitation and a
reduced population of the final target state, can also be used to avoid potential ionization.
Perhaps the most challenging aspect of the proposed scheme is the sinc function pulse envelope,
however, the ability to shape optical pulses is a rapidly developing field of
research~\citep{Goswami:PhysRep374:385}.

Our simulations utilized laser parameters well within the current capabilities of laser technology,
with calculations based on robust, first principles methodologies geared toward high accuracy, which
provide detailed information on the energy level structure and electric tensor transition moment
matrix elements. More sophisticated coherent control schemes could thus be developed to prepare
molecules of arbitrary structure in specific, highly excited rotational states. Extending this
approach to larger or heavier systems will pose challenges, most notably the increased density of
rotational states at lower energies, which makes it easier to populate undesired energy levels and
select the wrong rotational pathway, producing ``chaotic'' dynamics; this can be mitigated using
state-selection before applying the control pulses~\cite{Chang:IRPC34:557}. Despite these
difficulties, the ability to force stable rotation about different molecular axes and to change the
angular momentum alignment of a molecule would be highly advantageous in molecule-molecule or
molecule-surface scattering experiments, particularly with superrotors which display novel
properties and contain large amounts of rotational energy.

\begin{acknowledgments}
   This work has been supported by the \emph{Deutsche Forschungsgemeinschaft} (DFG) through the
   excellence cluster ``The Hamburg Center for Ultrafast Imaging -- Structure, Dynamics and Control
   of Matter at the Atomic Scale'' (CUI, EXC1074) and the priority program 1840 ``Quantum Dynamics
   in Tailored Intense Fields'' (QUTIF, KU1527/3), by the European Research Council under the
   European Union's Seventh Framework Programme (FP7/2007-2013) through the Consolidator Grant
   COMOTION (ERC-614507-Küpper), by the Helmholtz Gemeinschaft through the ``Impuls- und
   Vernetzungsfond'', and by the COST action MOLIM (CM1405).
\end{acknowledgments}

\bigskip
\def\bibsection{}
\centerline{\textbf{\textsc{\large references}}}
\medskip
\bibliographystyle{achemso}
\bibliography{string,cmi}

\providecommand{\refin}[1]{\\ \textbf{Referenced in:} #1}
\begin{thebibliography}{10}

\bibitem{Lemeshko:MP111:1648}
Lemeshko,~M.;\ \ Krems,~R.~V.;\ \ Doyle,~J.~M.;\ \ Kais,~S. Manipulation of
  molecules with electromagnetic fields. \textit{Mol.\ Phys.} \textbf{2013,}
  \textsl{111,} 1648-1682.

\bibitem{Stapelfeldt:RMP75:543}
Stapelfeldt,~H.;\ \ Seideman,~T. Colloquium: Aligning molecules with strong
  laser pulses. \textit{Rev.\ Mod.\ Phys.} \textbf{2003,} \textsl{75,} 543-557.

\bibitem{Holmegaard:PRL102:023001}
Holmegaard,~L.;\ \ Nielsen,~J.~H.;\ \ Nevo,~I.;\ \ Stapelfeldt,~H.;\ \
  Filsinger,~F.;\ \ K{\"u}pper,~J.;\ \ Meijer,~G. Laser-induced alignment and
  orientation of quantum-state-selected large molecules. \textit{Phys.\ Rev.\
  Lett.} \textbf{2009,} \textsl{102,} 023001.

\bibitem{Ghafur:NatPhys5:289}
Ghafur,~O.;\ \ Rouz\'ee,~A.;\ \ Gijsbertsen,~A.;\ \ Siu,~W.~K.;\ \ Stolte,~S.;\
  \ Vrakking,~M. J.~J. Impulsive orientation and alignment of
  quantum-state-selected {NO} molecules. \textit{Nat. Phys.} \textbf{2009,}
  \textsl{5,} 289-293.

\bibitem{Fleischer:NJP11:105039}
Fleischer,~S.;\ \ Khodorkovsky,~Y.;\ \ Prior,~Y.;\ \ Averbukh,~I.~S.
  Controlling the sense of molecular rotation. \textit{New J.\ Phys.}
  \textbf{2009,} \textsl{11,} 105039.

\bibitem{Kitano:PRL103:223002}
Kitano,~K.;\ \ Hasegawa,~H.;\ \ Ohshima,~Y. Ultrafast angular momentum
  orientation by linearly polarized laser fields. \textit{Phys.\ Rev.\ Lett.}
  \textbf{2009,} \textsl{103,} 223002.

\bibitem{Zhdanovich:PRL107:243004}
Zhdanovich,~S.;\ \ Milner,~A.~A.;\ \ Bloomquist,~C.;\ \ Flo{\ss},~J.;\ \
  Sh.~Averbukh,~I.;\ \ Hepburn,~J.~W.;\ \ Milner,~V. Control of molecular
  rotation with a chiral train of ultrashort pulses. \textit{Phys.\ Rev.\ Lett.}
  \textbf{2011,} \textsl{107,} 243004.

\bibitem{Ohshima:IRPC29:619}
Ohshima,~Y.;\ \ Hasegawa,~H. Coherent rotational excitation by intense
  nonresonant laser fields. \textit{Int.\ Rev.\ Phys.\ Chem.} \textbf{2010,}
  \textsl{29,} 619-663.

\bibitem{Milner:ACP159:395}
Milner,~V.;\ \ Hepburn,~J.~W. Laser control of ultrafast molecular rotation.
  \textit{Adv.\ Chem.\ Phys.} \textbf{2016,} \textsl{159,} 395-412.

\bibitem{Karczmarek:PRL82:3420}
Karczmarek,~J.;\ \ Wright,~J.;\ \ Corkum,~P.;\ \ Ivanov,~M. Optical centrifuge
  for molecules. \textit{Phys.\ Rev.\ Lett.} \textbf{1999,} \textsl{82,}
  3420--3423.

\bibitem{Villeneuve:PRL85:542}
Villeneuve,~D.~M.;\ \ Aseyev,~S.~A.;\ \ Dietrich,~P.;\ \ Spanner,~M.;\ \
  Ivanov,~M.~Y.;\ \ Corkum,~P.~B. Forced molecular rotation in an optical
  centrifuge. \textit{Phys.\ Rev.\ Lett.} \textbf{2000,} \textsl{85,} 542-545.

\bibitem{Yuan:PNAS108:6872}
Yuan,~L.;\ \ Teitelbaum,~S.~W.;\ \ Robinson,~A.;\ \ Mullin,~A.~S. Dynamics of
  molecules in extreme rotational states. \textit{Proc. Natl. Acad. Sci. U. S.
  A.} \textbf{2011,} \textsl{108,} 6872-6877.

\bibitem{Milner:PRL113:043005}
Milner,~A.~A.;\ \ Korobenko,~A.;\ \ Hepburn,~J.~W.;\ \ Milner,~V. Effects of
  ultrafast molecular rotation on collisional decoherence. \textit{Phys.\ Rev.\
  Lett.} \textbf{2014,} \textsl{113,} 043005.

\bibitem{Khodorkovsky:NatComm6:7791}
Khodorkovsky,~Y.;\ \ Steinitz,~U.;\ \ Hartmann,~J.-M.;\ \ Sh.~Averbukh,~I.
  Collisional dynamics in a gas of molecular super-rotors. \textit{Nat. Commun.}
  \textbf{2015,} \textsl{6,} 7791.

\bibitem{Korobenko:PRL116:183001}
Korobenko,~A.;\ \ Milner,~V. Adiabatic field-free alignment of asymmetric top
  molecules with an optical centrifuge. \textit{Phys.\ Rev.\ Lett.}
  \textbf{2016,} \textsl{116,} 183001.

\bibitem{Omiste:PRA97:023407}
Omiste,~J.~J. Theoretical study of asymmetric super-rotors: Alignment and
  orientation. \textit{Phys.\ Rev.\ A} \textbf{2018,} \textsl{97,} 023407.

\bibitem{Tutunnikov:JPCL9:1105}
Tutunnikov,~I.;\ \ Gershnabel,~E.;\ \ Gold,~S.;\ \ Sh.~Averbukh,~I. Selective
  orientation of chiral molecules by laser fields with twisted polarization.
  \textit{J.\ Phys.\ Chem.\ Lett.} \textbf{2018,} \textsl{9,} 1105--1111.

\bibitem{Godsi:NatComm8:15357}
Godsi,~O.;\ \ Corem,~G.;\ \ Alkoby,~Y.;\ \ Cantin,~J.~T.;\ \ Krems,~R.~V.;\ \
  Somers,~M.~F.;\ \ Meyer,~J.;\ \ Kroes,~G.-J.;\ \ Maniv,~T.;\ \
  Alexandrowicz,~G. A general method for controlling and resolving rotational
  orientation of molecules in molecule-surface collisions. \textit{Nat. Commun.}
  \textbf{2017,} \textsl{8,} 15357.

\bibitem{Yurchenko:JMS245:126}
Yurchenko,~S.~N.;\ \ Thiel,~W.;\ \ Jensen,~P. Theoretical {ROVibrational}
  energies ({TROVE}): A robust numerical approach to the calculation of
  rovibrational energies for polyatomic molecules. \textit{J.\ Mol.\ Spectrosc.}
  \textbf{2007,} \textsl{245,} 126--140.

\bibitem{Yachmenev:JCP143:014105}
Yachmenev,~A.;\ \ Yurchenko,~S.~N. Automatic differentiation method for
  numerical construction of the rotational-vibrational {H}amiltonian as a power
  series in the curvilinear internal coordinates using the {E}ckart frame.
  \textit{J.\ Chem.\ Phys.} \textbf{2015,} \textsl{143,} 014105.

\bibitem{Yurchenko:JCTC13:4368}
Yurchenko,~S.~N.;\ \ Yachmenev,~A.;\ \ Ovsyannikov,~R.~I. Symmetry adapted
  ro-vibrational basis functions for variational nuclear motion calculations:
  {TROVE} approach. \textit{J.\ Chem.\ Theory\ Comput.} \textbf{2017,}
  \textsl{13,} 4368.

\bibitem{Azzam:MNRAS460:4063}
Azzam,~A. A.~A.;\ \ Tennyson,~J.;\ \ Yurchenko,~S.~N.;\ \ Naumenko,~O.~V.
  {ExoMol} molecular line lists {\textendash} {XVI}. the
  rotation{\textendash}vibration spectrum of hot {H$_2$S}. \textit{Mon. Not. R.
  Astron.\ Soc.} \textbf{2016,} \textsl{460,} 4063--4074.

\bibitem{Aidas:WIRE4:269}
Aidas,~K. \textit{et al.}\  The {D}alton quantum chemistry program system.
  \textit{WIREs Comput.~Mol.~Sci.} \textbf{2014,} \textsl{4,} 269--284.

\bibitem{Owens:JCP148:124102}
Owens,~A.;\ \ Yachmenev,~A. {RichMol}: A general variational approach for
  rovibrational molecular dynamics in external electric fields. \textit{J.\
  Chem.\ Phys.} \textbf{2018,} \textsl{148,} 124102.

\bibitem{Trippel:MP111:1738}
Trippel,~S.;\ \ Mullins,~T.;\ \ M{\"u}ller,~N. L.~M.;\ \ Kienitz,~J.~S.;\ \
  D{\l}ugo{\l}\k{e}cki,~K.;\ \ K{\"u}pper,~J. Strongly aligned and oriented
  molecular samples at a {kHz} repetition rate. \textit{Mol.\ Phys.}
  \textbf{2013,} \textsl{111,} 1738.

\bibitem{Pyka:MolPhys70:547}
Pyka,~J. Clustering of rovibrational energy levels in the highly excited
  {H$_2$S} molecule. \textit{Mol.\ Phys.} \textbf{1990,} \textsl{70,} 547--561.

\bibitem{Kozin:JMolSpec163:483}
Kozin,~I.~N.;\ \ Jensen,~P. Fourfold clusters of rovibrational energy levels
  for {H$_2$S} studied with a potential energy surface derived from experiment.
  \textit{J.\ Mol.\ Spectrosc.} \textbf{1994,} \textsl{163,} 483--509.

\bibitem{Jensen:WCMS2:494}
Jensen,~P. Local modes in vibration-rotation spectroscopy. \textit{WIREs Comput.
  Mol. Sci.} \textbf{2012,} \textsl{2,} 494-512.

\bibitem{Bunker:JMolSpec228:640}
Bunker,~P.~R.;\ \ Jensen,~P. Chirality in rotational energy level clusters.
  \textit{J.\ Mol.\ Spectrosc.} \textbf{2004,} \textsl{228,} 640-644.

\bibitem{Owens:RIC:inprep}
Owens,~A.;\ \ Yachmenev,~A.;\ \ Yurchenko,~S.~N.;\ \ K\"{u}pper,~J. Climbing
  the rotational ladder to chirality. \textit{submitted} \textbf{2018,} \textit{arXiv:1802.07803 [physics.chem-ph]}.

\bibitem{Goswami:PhysRep374:385}
Goswami,~D. Optical pulse shaping approaches to coherent control. \textit{Phys.\
  Rep.} \textbf{2002,} \textsl{374,} 385-481.

\bibitem{Chang:IRPC34:557}
Chang,~Y.-P.;\ \ Horke,~D.~A.;\ \ Trippel,~S.;\ \ K{\"u}pper,~J.
  Spatially-controlled complex molecules and their applications. \textit{Int.\
  Rev.\ Phys.\ Chem.} \textbf{2015,} \textsl{34,} 557-590.

\end{thebibliography}
\onecolumngrid
\end{document}